\documentclass[sigconf,natbib=true]{acmart}

\author{Stan Fris}

\email{s.c.j.fris@uva.nl}

\affiliation{\institution{University of Amsterdam}\country{The Netherlands}}

\author{Jan Hutter}
\email{jan.hutter@student.uva.nl}

\affiliation{\institution{University of Amsterdam}\country{The Netherlands}}

\author{Jan Henrik Bertrand}
\email{jan.henrik.bertrand@student.uva.nl}

\affiliation{\institution{University of Amsterdam}\country{The Netherlands}}

\author{Simon Lupart}
\email{s.c.lupart@uva.nl}

\affiliation{\institution{University of Amsterdam}\country{The Netherlands}}

\author{Mohammad Aliannejadi}
\email{m.aliannejadi@uva.nl}

\affiliation{\institution{University of Amsterdam}\country{The Netherlands}}
\usepackage[capitalise]{cleveref}
\usepackage{booktabs}
\usepackage{multirow}
\crefname{enumi}{}{}
\usepackage{amsmath}
\usepackage{mathtools}

\usepackage[inline]{enumitem}
\usepackage{siunitx}
\usepackage{float}
\usepackage[printonlyused]{acronym}

\usepackage{xspace}
\newcommand{\disco}{DiSCo\xspace}

\newcommand{\personalnote}[1]{#1}

\newcommand{\ourmethod}[1]{\personalnote{\textcolor{gray}{NAME OF OUR METHOD}}}

\setcopyright{none}

\acmDOI{}
\acmISBN{}

\copyrightyear{2026}
\acmYear{2026}
\acmConference[SCAI]{SCAI Workshop at SIGIR '26}{July 20--24, 2026}{Melbourne, Naarm, Australia}
\acmBooktitle{SCAI Workshop at SIGIR '26, July 20--24, 2026, Melbourne, Naarm, Australia}


\begin{abstract}
    Conversational Search (CS) considers retrieval of relevant documents based on conversational context. 
    Large Language Models (LLMs) have significantly enhanced CS by enabling effective query rewriting.
    However, employing LLMs during inference poses efficiency challenges. A method to balance effectiveness and efficiency is the use of knowledge distillation from LLM-based query-rewriting. Recent work applies the Kullback–Leibler Divergence (KLD) for distillation, relaxing on the alignment with the teacher signal compared to previous methods. 
    
    Despite these gains, several aspects of KLD-based distillation for conversational search remain understudied, and we investigate them in this work. Prior work in related fields suggests that adding a contrastive loss to the KLD objective can improve performance; we confirm this and observe significant gains in precision oriented ranking metrics. We also find that contrastive sampling strategies for the KLD loss have a non-trivial impact and must be chosen carefully. Although theory suggests that more samples improve the KLD estimate, experiments show diminishing returns on the number of used samples. Finally, we address the phenomenon of decreased sparsity in longer conversations, which limits computational efficiency across sparse retrieval methods. We find that the representations from the model distilled with the KLD loss can be strongly regularized with a regularization loss, substantially improving sparsity and inference efficiency without significantly harming retrieval effectiveness. We achieve a $2\times$ decrease in FLOPS on TopiOCQA with negligible loss in effectiveness ($\leq 2\%$ drop on Recall@100). 
    Our results provide insights into distillation objectives for learned sparse conversational retrievers and offer practical guidelines for improving effectiveness and efficiency in first-stage retrieval. 
\end{abstract}

\begin{document}

\title{Improving the Efficiency and Effectiveness of LLM Knowledge Distillation for Conversational Search}

\maketitle

\acrodef{LLM}{Large Language Model}
\acrodef{MSE}{Mean-Squared Error}
\acrodef{KLD}{Kullback-Leibler Divergence}
\acrodef{LSR}{Learned Sparse Retrieval}
\acrodef{CQR}{Conversational Query Rewriting}
\acrodef{CS}{Conversational\- Search}

\section{Introduction} \label{sec:introduction}

First-stage retrieval is an important part of search systems, aiming to maximize recall~\cite{chen_efficient_2017, matveeva_high_2006} before a second-stage ranker yields the final ranking order. 
Given that the first-stage retrieval is done on a much larger collection (order of millions), the main focus is to design efficient algorithms while ensuring maximum recall~\cite{xiao_beyond_2023,lassance_efficiency_2022}. An effective approach for first-stage retrieval is the use of sparse retrievers, which leverage inverted index structures for efficient large-scale search \cite{formal_splade_2021, lassance_efficiency_2022, lupart_disco_2025, arabzadeh_predicting_2021, nguyen2023unified}.

\ac{CS}~\cite{mo_survey_2025} is a task that centers on modeling the evolution of users' information goals over the course of the interaction. A key challenge is to remain accurate when more turns are added to the conversation, and to separate relevant from irrelevant information when contextualizing. An approach that has been shown to be effective is the use of Large Language Models (LLMs) for conversation rewriting~\cite{yu_few-shot_2021, lupart_disco_2025, hai_cosplade_2024, mao_learning_2023, wu_conqrr_2022}. However, LLM inference is a computationally expensive operation, increasing latency and cost. An approach to balance this cost is the use of knowledge distillation for learning representations~\cite{hai_cosplade_2024, mao_learning_2023}. Recent research has shown improvements in representations when using student-teacher setup for learning to represent rewritten conversations in sparse retrieval~\cite{hai_cosplade_2024, mao_learning_2023} These approaches make use of a \ac{MSE} loss to make the student query representation approach the LLM-rewritten teacher query representation, and apply these to perform first-stage retrieval with sparse retrievers.

Research by~\citet{lupart_disco_2025} suggests that this approach can be further improved by using a \ac{KLD} loss \textit{between the similarity scores} generated by the student and teacher models~\cite{kullback_information_1951}. This method provides more freedom in the representation space, and has state-of-the-art performance in the CS domain. In this work, we theoretically and empirically investigate several components of the use of the \ac{KLD} for knowledge distillation in \ac{CS}, including the loss function, sampling method, and regularization. 

To the best of our knowledge, previous work only considers the use of the \ac{KLD} in \ac{CS} as the sole loss objective~\cite{lupart_disco_2025}. Related work has shown that adding a contrastive component can lead to improved performance in ranking settings~\cite{yang_adaptive_2024, yang_balanced_2023, yu_few-shot_2021, lassance2024splade}.
We study to what extent the \ac{KLD} loss can be combined with a contrastive loss to improve model performance. Furthermore, the number of samples used in the \ac{KLD} loss remains underexplored. When applied in \ac{CS}, the \ac{KLD} is based on contrastive sampling for the \ac{KLD}, as in retrieval settings there are target (positive) and non-target (negative) samples. This suggests a balance in sampling, where the positive sample needs to be represented sufficiently~\cite{yang_balanced_2023}. We theoretically and experimentally investigate this setting, looking at how different sampling affects the \ac{KLD}, as well as downstream performance metrics. 

Finally, we investigate sparsity and inference speed: prior work applying the \ac{KLD} objective to \ac{CS} uses regularization similar to standard sparse retrieval methods and observes deteriorating sparsity and performance for longer sequences~\cite{lupart_disco_2025, formal_splade_2021}. The \ac{KLD} based objective is a relaxation compared to MSE distillation objectives, suggesting that regularization can be applied to a larger extent. 
We investigate the effect of applying large amounts of regularization, looking at sparsity and performance. 

To summarize, we propose analyzing several topics to improve effectiveness and efficiency of \ac{KLD} distillation in \ac{CS}. Our work tackles the following three research questions:

\begin{enumerate}[label=\textbf{RQ\arabic*:},leftmargin=*, align=left,labelwidth=3em, labelsep=0.5em, ref={\textbf{RQ\arabic*}}]
    \item \label{rq1} To what extent can a KL divergence distillation objective be combined with a contrastive InfoNCE for improved performance in conversational search?
    \item \label{rq2} How does the balance of positive and negative samples used in the KL Divergence objective affect distillation behavior and downstream retrieval performance?
    \item \label{rq3} How does increasing regularization in DiSCo influence the trade-off between sparsity, inference efficiency, and retrieval effectiveness, particularly for longer conversational queries?
\end{enumerate}

Our contributions are the following: We first show that the use of a contrastive loss can significantly improve performance on ranking metrics when applied in minor amounts (\cref{rq1}). We then demonstrate that while the alignment with the objective increases as more samples are used, when many samples are used the positive sample is underrepresented, leading to diminishing returns in terms of performance metrics (\cref{rq2}). Finally, we find that large amounts of regularization can be applied effectively, keeping representations remarkably sparse for longer conversations, without significantly affecting performance (\cref{rq3}).

\section{Methodology} \label{sec:Method}

\subsection{Objective}
\label{subsec:method:objective}
\citeauthor{lupart_disco_2025} consider the use of KLD over the similarity scores for one positive and 16 negative samples, effectively using this as both a distillation and contrastive incentive. While using this setup might intuitively resemble a contrastive incentive to a certain extent, the objective behind the \ac{KLD} is the alignment of distributions rather than contrasting positive and negative examples~\cite{kullback_information_1951}. This approach does not allow for a fine-grained control of the contrastive objective in combination with the distillation objective. Several related works report improvements in precision-based metrics when adding a weighted contrastive component to the objective ~\cite{yang_adaptive_2024, yang_balanced_2023}. 

We propose a weighted combination of a \ac{KLD} based distillation loss and the InfoNCE loss as a controllable contrastive component:

\begin{equation}
\begin{split}
Loss = \lambda \cdot InfoNCE_{sim}  + (1-\lambda) \cdot KLD 
\end{split}
\label{Contrastive_objective}
\end{equation}

where \(\lambda \in [0,1]\) acts as a weighting parameter to balance the contrastive and the distillation incentive, \textit{KLD} is the KLD objective specified in~\citet{lupart_disco_2025}, and $infoNCE_{sim}$ is the infoNCE~\cite{oord2018representation} loss of the similarity scores. 

\subsection{Ratio of Negatives and Positives in Sampling}
\label{subsec:method:ratio_neg_pos}
In addition to improving the objective, using a larger number of positive samples could allow for a more representative distribution and thus a more informed loss. This could enable a faster training convergence through enhanced training signal. Therefore, there remains a research gap in investigating to what extent alternate sampling strategies can be used with the KLD.
Specifically, we can show that utilizing more samples from the teacher inside the KLD loss will lead to a better approximation of the teacher similarity scores distribution. Starting from the teacher similarity scores $T_\tau(i)$ and student similarity scores $S_\tau(i)$:

\begin{equation}
T_\tau(i) = \frac{\exp(s_i^{(T)}/\tau)}{\sum_{j=1}^n \exp(s_j^{(T)}/\tau)}, 
\qquad
S_\tau(i) = \frac{\exp(s_i^{(S)}/\tau)}{\sum_{j=1}^n \exp(s_j^{(S)}/\tau)}.
\end{equation}
\vspace*{-0.2cm}

The KL divergence from teacher to student, written as an expectation, is: 
\vspace*{-0.1cm}
\[
D_{\mathrm{KL}}(T_\tau \,\|\, S_\tau)
= \mathbb{E}_{i \sim T_\tau}\left[\log \frac{T_\tau(i)}{S_\tau(i)}\right]
= \sum_{i=1}^n T_\tau(i)\,\log\frac{T_\tau(i)}{S_\tau(i)}.
\]
\vspace*{-0.3cm}

Here we can see that from the law of large numbers, increasing the number of samples drawn from the teacher model yields a more accurate approximation of the true expectation in the \ac{KLD}.

Furthermore, adding more samples from high-probability areas will lead to a more effective representation of the teacher distribution, as these samples are underrepresented in the training objective. This is similar to the intuition of importance sampling~\cite{tokdar_importance_2010}, where the sampling distribution is adjusted to place greater weight on regions that contribute most to the target expectation, thereby reducing variance and improving the accuracy of the estimator.

\subsection{Regularization of Relaxed Objective}

An important motivation for the use of the \ac{KLD} in knowledge distillation is its relaxation compared to contrastive or MSE-based methods. In previous work by \citet{lupart_disco_2025}, the analysis shows that, despite this relaxation, the resulting representations exhibit sparsity comparable to that obtained with LLM-based query rewriting or MSE-based distillation. We argue that this is because an insufficient amount of regularization was applied on the sparse representations. The \ac{KLD} objective should allow for more general representations as the loss objective does not consider individual representations, therefore, we hypothesize that more regularization can be applied before diminishing returns are seen. 

We investigate this by applying $L_1$~\cite{10.1145/3269206.3271800} to query representations and FLOPS~\cite{paria2020minimizing} to document representations, aligning with previous work~\cite{formal_splade_2021,lupart_disco_2025}. We measure efficiency using the FLOPs metric~\cite{paria2020minimizing, formal2021spladev2}, which estimates the expected number of non-zero floating-point operations incurred during query–document matching in an inverted-index retrieval setting:
\begin{equation}
\mathrm{FLOPs}
= \mathbb{E}_{q,d}
\left[
\sum_{j \in V}
p^{(q)}_{j}\, p^{(d)}_{j}
\right],
\end{equation}
Where $p_j^{(q)}$ and $p_j^{(p)}$ are the activation probabilities of the $j$th token for the query and document representations respectively.
Finally, we investigate how regularization affects conversations with a larger number of turns, as previous work showed that sparsity commonly decreases with conversation length~\cite{lupart_disco_2025}.

\section{Experiments}
In this section, we validate our claims with respect to the objective, sampling and regularization. To the best of our knowledge, the only existing implementation of a KLD loss for knowledge distillation in conversational search is \disco \cite{lupart_disco_2025}. We apply our experiments to their implementations, and use SPLADE++~\cite{formal_distillation_2022} for all experiments. In preliminary experiments, we found that extending training from 5 to 7 epochs led to improved convergence, therefore, we use this for all our experiments. We utilized the same hyperparameters as the original authors. Our results are shown for the TopiOCQA~\cite{noauthor_topiocqa_nodate} dataset, a large conversational dataset derived from Natural Questions, which is a widely used benchmark~\cite{lupart_disco_2025, lassance_efficiency_2022, hai_cosplade_2024, formal_splade_2021, mao_learning_2023, mo_aligning_2024, yu_few-shot_2021}.

\subsection{RQ1: Dual Loss Objective}
Our analysis in~\cref{subsec:method:objective} indicated that Knowledge Distillation can potentially be improved by adding a contrastive component, motivated by related work \citep{yang_balanced_2023, yang_adaptive_2024}. We propose adding an InfoNCE loss, and investigate how this affects model performance.

We investigate different proportions of InfoNCE following~\cref{Contrastive_objective}, with $\lambda \in \{0.05, 0.10, 0.20\}$, where each number indicates the proportion of the loss represented by the InfoNCE objective.
In~\Cref{tab:disco_with_contrastive} we show the results of training our model with varying amounts of contrastive loss.  We observe that using an InfoNCE component of 10\% or 20\% leads to significantly improved performance on ranking metrics compared to DiSCo. For recall-based metrics, we see that Recall@100 does decrease slightly, although we do not find that these decreases are significant. Recall@10 results are similar across models, with a slight improvement for a 5\% InfoNCE component. 

These results confirm the theoretical motivation that the addition of a ranking loss improves the ranking capacity, increasing performance in terms of MRR and nDCG@3. Additionally, weS observe that this can lead to higher Recall@k scores for lower k values, possibly due to the improved ranking of individual items, leading to improved recall with small cutoffs. Our results imply that in settings where ranking performance is relevant, adding InfoNCE loss as a component improves model effectiveness. In cases where only recall is relevant, a contrastive loss is not necessary, but does not significantly deteriorate performance.

\begin{table}[h!]
\caption{Results on TopiOCQA after adding a weighted contrastive loss (InfoNCE) to the KL divergence. The weight of DiSCo and InfoNCE sum to 1. $\star$ is for significantly better and $\dagger$ for significantly worse results under a Bonferroni-corrected paired t-test with a \(95\%\) confidence level.}
\vspace{-0.3cm}
\centering
\resizebox{\linewidth}{!}{
\begin{tabular}{lcccc}
\toprule
\textbf{Loss} & \textbf{MRR} & \textbf{R@10} & \textbf{R@100} & \textbf{nDCG@3} \\
\midrule
DiSCo & 0.409 & 0.676 & \textbf{0.879} & 0.396 \\
DiSCo + 0.05 InfoNCE & 0.414 & \textbf{0.687} & 0.875 & 0.400 \\
DiSCo + 0.10 InfoNCE & 0.421$^\star$ & 0.673 & 0.877 & 0.408$^\star$ \\
DiSCo + 0.20 InfoNCE & \textbf{0.424$^\star$} & 0.676 & 0.874 & \textbf{0.412$^\star$} \\
\bottomrule
\end{tabular}
}
\vspace*{-0.5cm}
\label{tab:disco_with_contrastive}
\end{table}

\section{RQ2: Negative Sampling in the KL Divergence}
As proven in~\cref{subsec:method:ratio_neg_pos}, the number of samples drawn from the teacher models should yield a more accurate approximation of the true expectation in the KLD. To evaluate alignment, we visualize the sample distributions on the test set to compare them to the teacher model, as shown in~\cref{fig:recall_across_sequence_lens}. We investigated different numbers of negative samples for the KLD. Here, we observe that the teacher and student models are generally not aligned. Although the models might be able to effectively learn representations leading to similar or better performance on ranking metrics, we don't find that representations exactly match those from the teacher model. However, the shape of the frequency distribution varies with the number of samples; in particular, when only a single negative sample is considered, differences between individual curves are minimal.

\begin{figure}[h!]\includegraphics[width=\linewidth, trim=0 20 0 10]{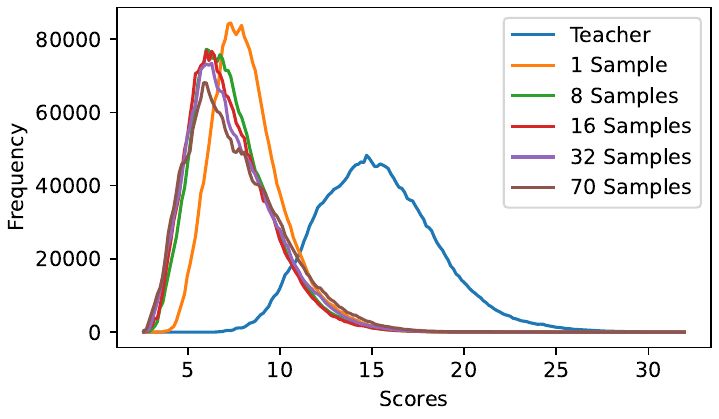}
        \caption{Distributions over the predicted similarity score for documents in the TopiOCQA test set, with teacher distribution (SPLADE MistralQR) for reference.}
    \label{fig:recall_across_sequence_lens}
    \vspace*{-0.4cm}
\end{figure}

We also consider increasing the number of negative samples for the KLD to improve performance. The results of these experiments, as well as the KLD values, are shown in~\autoref{tab:disco_with_different_sample_amounts}. Contrary to our hypothesized improvements in performance for increased sample amounts, we observe that performance does not increase with a larger sample size in practice. Sampling 16 negatives leads to the best results, with performance being significantly better than other models in most settings. Looking at the KL divergence, we observe this does become lower when using more samples, indicating that distributions do move closer when using more samples, confirming our intuition from~\cref{subsec:method:ratio_neg_pos}. 

\begin{table}[h!]
\caption{Results on TopiOCQA with different numbers of negatives in the KL divergence. There is always one positive. $\star$ is for significantly better and $\dagger$ for significantly worse results under a Bonferroni-corrected paired t-test with a \(95\%\) confidence level.}
\vspace{-0.3cm}
\centering
\resizebox{\linewidth}{!}{
\begin{tabular}{lcccccc}
\toprule
\textbf{Sample Amount} & \textbf{MRR} & \textbf{R@10} & \textbf{R@100} & \textbf{nDCG@3} & \textbf{D$_{KL}$} \\
\midrule
1  & 0.348$^\dagger$ & 0.613$^\dagger$ & 0.859$^\dagger$ & 0.326$^\dagger$ & 3.501 \\
8  & 0.394$^\dagger$ & 0.657$^\dagger$ & 0.874 & 0.381$^\dagger$ & 3.861 \\
16 & 0.409 & \textbf{0.676} & \textbf{0.879} & \textbf{0.396} & 3.698 \\
32 & \textbf{0.410} & 0.662$^\dagger$ & 0.874 & 0.395 & 3.592 \\
70 & 0.395$^\dagger$ & 0.655$^\dagger$ & 0.860$^\dagger$ & 0.381$^\dagger$ & 3.221 \\
\bottomrule
\end{tabular}}
\vspace*{-0.3cm}
\label{tab:disco_with_different_sample_amounts}
\end{table}

One explanation is that the positive sample is under-represented using a higher number of samples. In the \ac{KLD} loss, the positive sample is included in the negative sample distribution and given the score of the highest sample. By including more samples, the influence of the positive sample on the distribution and therefore the loss decreases. This means that gradient updates will be less focused on the positive sample. This can lead to less fine-grained learning for this sample and ultimately decreased performance. This explanation aligns with findings by~\citet{yang_adaptive_2024}, who investigate weighting of the \ac{KLD} in ranking objectives, and adds further nuance: We find that even in settings where the \ac{KLD} is used as a loss in the indirect task of similarity scoring, the relative representation of the positive sample has an important role.

\subsection{RQ3: Sparsity and Inference Speed}
In this section, we investigate to what extent regularization can improve sparsity and therefore inference speed in a model distilled with the KLD loss. We apply five regularization settings, regularization including the weights used by DiSCo \cite{lupart_disco_2025} and SPLADE \cite{formal_splade_2021}: None ($\lambda_d=\lambda_q=0$), DiSCo/SPLADE ($\lambda_d=5\times10^{-4}$, $\lambda_q=10^{-3}$), High ($\lambda_d=10^{-3}$, $\lambda_q=5\times10^{-3}$), Higher ($\lambda_d=10^{-2}$, $\lambda_q=5\times10^{-2}$), and Highest ($\lambda_d=5\times10^{-2}$, $\lambda_q=10^{-1}$). In~\cref{tab:regularization}, we show the results of adding increasing amounts of regularization. 
Here, we observe that with increasing regularization, there is little change in performance while substantially improving the FLOPS. For the ``High'' setting, only Recall\@10 shows a statistically significant degradation, while all other metrics remain within the significance bounds. For the higher $\lambda$ setting, all values are within the same threshold, while reaching a FLOPS score that is two times lower. 
\begin{table}[h]
\caption{Results on TopiOCQA after adding regularization to DiSCo, where SPLADE indicates the Regularization applied for SPLADE. $\star$ indicates significantly better and $\dagger$ significantly worse results under a Bonferroni-corrected paired t-test with a \(95\%\) confidence level.}
\vspace{-0.3cm}
\resizebox{\linewidth}{!}{
\centering
\begin{tabular}{lrrrrr}
\toprule
\textbf{Loss / Setting} & \textbf{MRR} & \textbf{R@10} & \textbf{R@100} & \textbf{nDCG@3} & \textbf{FLOPS} \\
\midrule
None & 0.409 & \textbf{0.676} & 0.879 & 0.396 & 3.790 \\
DiSCo Reg. & 0.405 & 0.659$^\dagger$ & 0.877 & 0.391 & 3.140 \\
High & \textbf{0.412} & 0.667 & 0.872 & \textbf{0.398} & 1.840 \\
Higher & \textbf{0.412} & 0.662 & 0.859 & 0.394 & 1.370 \\
Highest & 0.385$^\dagger$ & 0.619$^\dagger$ & 0.825$^\dagger$ & 0.371$^\dagger$ & \textbf{0.470} \\
\bottomrule
\end{tabular}
}
\vspace*{-0.3cm}
\label{tab:regularization}
\end{table}

\begin{figure}[h!]
\includegraphics[width=0.95\linewidth, trim=0 50 0 15]{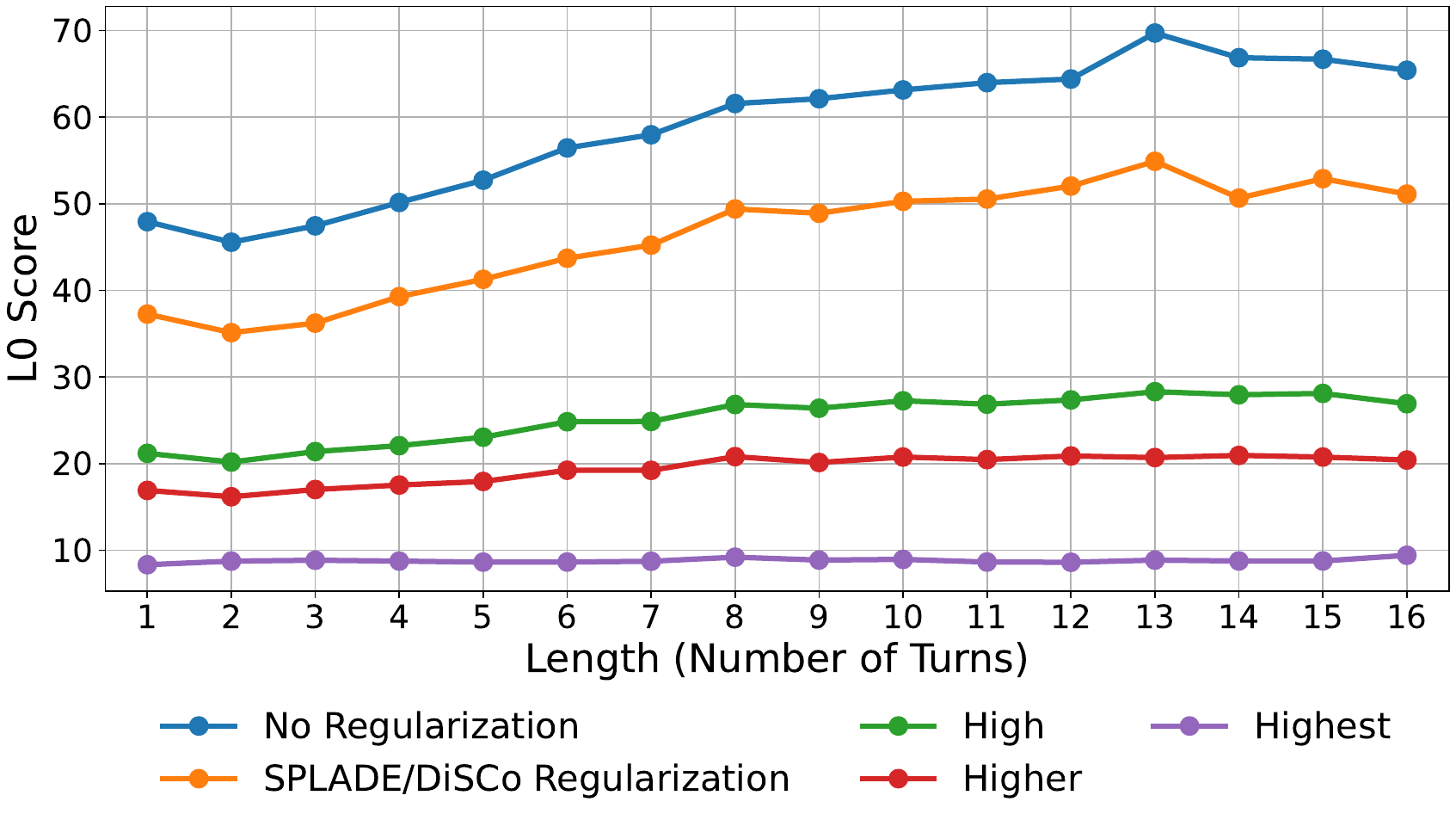}
    \caption{The L0 Score as a measure of sparsity for different conversation lengths (measured by the number of turns) across a range of regularization settings on TopiOCQA. Lower is better.}
    \label{fig:feature_separation_sampling}
    \vspace*{-0.5cm}
\end{figure}

\citet{lupart_disco_2025} show that the amount of activated embedding dimensions increases as the number of conversation turns increases. We investigate how this increase appears in our findings, and compare it with regularized versions. In~\cref{fig:feature_separation_sampling}, we show sparsity across conversations with different amounts of turns. When we compare the original DiSCo setting with versions where increased regularization was applied, we see a significant improvement in sparsity of representations. Furthermore, we find that while DiSCo becomes proportionally much sparser with increased conversation lengths compared to the higher $\lambda$ settings. 

We also investigate performance across different numbers of turns, as shown in~\cref{fig:recall_across_sequence_lens}. Here, we see that for all models, performance generally decreases as the number of conversation turns increases. This can be explained by the fact that increased-length conversations are generally more complex and require the model to encode more information. Looking at the performance of regularized methods compared to DiSCo we observe that performance for ``High'' and ``Higher'' regularization is very similar to DiSCo performance. This indicates that regularization can be applied effectively to both long and shorter conversations, as performance changes compared to a non-regularized version are small.  

Overall, we observe that regularization can be an effective method for the improvement of the inference speed of DiSCo. Regularization leads to especially large improvements for longer conversations without suffering from decreased performance over DiSCo.

\begin{figure}[h!]    \includegraphics[width=0.95\linewidth, trim=0 50 0 15]{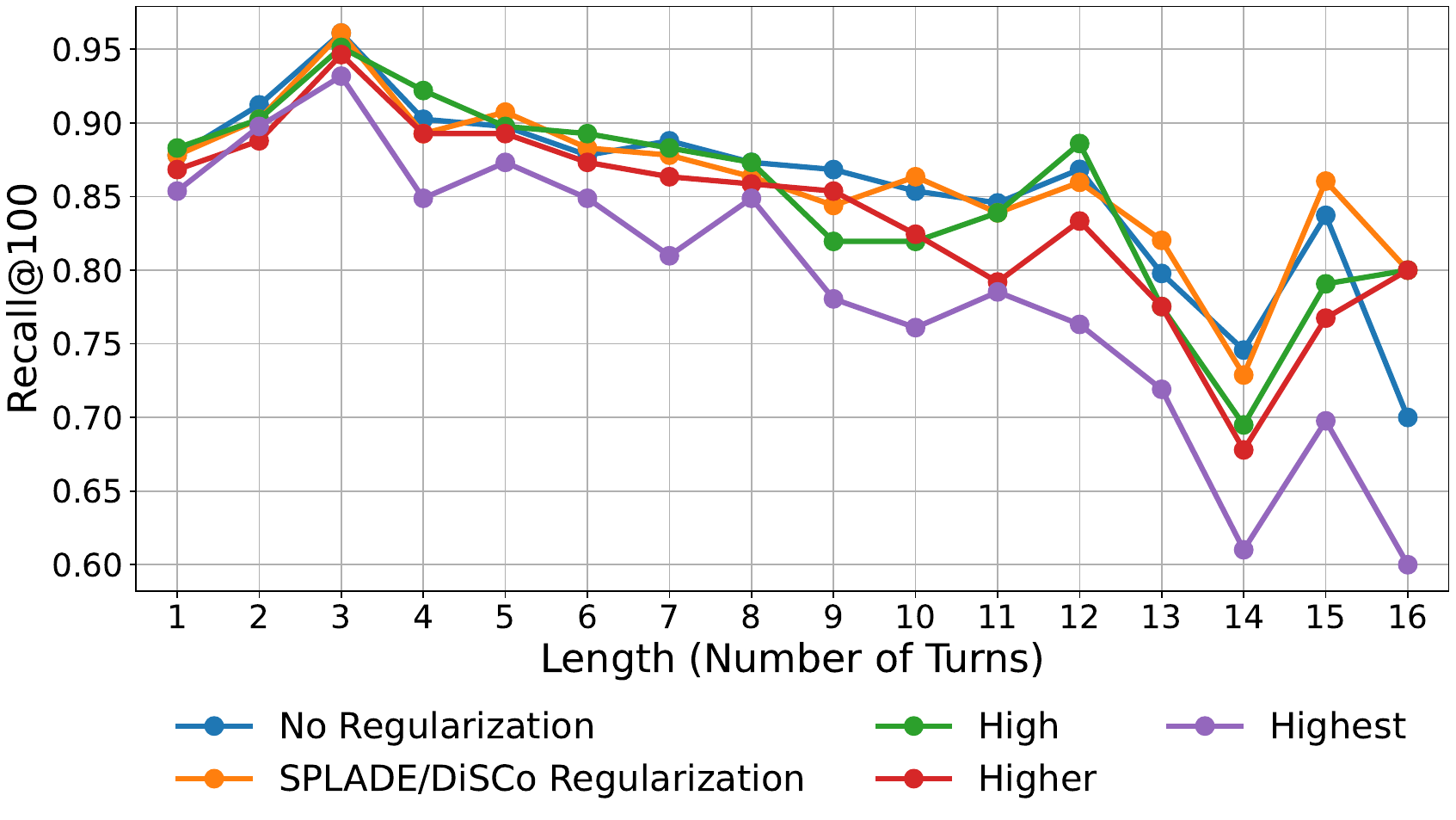}
    \caption{Recall@100 on TopiOCQA for different conversation lengths (measured by the number of turns) across a range of regularization settings. Higher is better.}
    \label{fig:recall_across_sequence_lens}
    \vspace*{-0.5cm}
\end{figure}

\label{sec:discussion}
\section{Conclusion}
\label{sec:conclusion}

In this work, we study components to improve the use of the KL Divergence as a contrastive objective for knowledge-distillation in Conversational Search. By suggesting the addition of a contrastive loss, we show that performance on ranking metrics can be improved when contrastive learning is added. We provide insight to practitioners by showing that while theory might suggest to include a large number of samples in the KLD, when using a contrastive objective a balancing of positive and negative samples is required. Finally, we investigate regularization, where we verify our claim that KLD based methods can be regularized significantly more. We find that we can use more regularization, without negative effects on recall, and we find that more regularization helps keep the representations remarkably sparse for longer conversations, providing a solution to a significant issue in conversational search. Our work addresses several open directions in Knowledge Distillation for Conversational Search, and provides actionable insights.

\newpage
\bibliographystyle{ACM-Reference-Format}
\bibliography{newreferences}

@inproceedings{formal_distillation_2022,
	address = {Madrid Spain},
	title = {From {Distillation} to {Hard} {Negative} {Sampling}: {Making} {Sparse} {Neural} {IR} {Models} {More} {Effective}},
	isbn = {978-1-4503-8732-3},
	shorttitle = {From {Distillation} to {Hard} {Negative} {Sampling}},
	url = {https://dl.acm.org/doi/10.1145/3477495.3531857},
	doi = {10.1145/3477495.3531857},
	language = {en},
	urldate = {2025-12-12},
	booktitle = {Proceedings of the 45th {International} {ACM} {SIGIR} {Conference} on {Research} and {Development} in {Information} {Retrieval}},
	publisher = {ACM},
	author = {Formal, Thibault and Lassance, Carlos and Piwowarski, Benjamin and Clinchant, Stéphane},
	month = jul,
	year = {2022},
	pages = {2353--2359},
}

@article{tokdar_importance_2010,
	title = {Importance sampling: a review},
	volume = {2},
	copyright = {Copyright © 2009 John Wiley \& Sons, Inc.},
	issn = {1939-0068},
	shorttitle = {Importance sampling},
	url = {https://onlinelibrary.wiley.com/doi/abs/10.1002/wics.56},
	doi = {10.1002/wics.56},
	language = {en},
	number = {1},
	urldate = {2025-11-27},
	journal = {WIREs Computational Statistics},
	author = {Tokdar, Surya T. and Kass, Robert E.},
	year = {2010},
	note = {\_eprint: https://wires.onlinelibrary.wiley.com/doi/pdf/10.1002/wics.56},
	pages = {54--60},
}

@misc{noauthor_topiocqa_nodate,
	title = {{TopiOCQA}: {Open}-domain {Conversational} {Question} {Answering} with {Topic} {Switching} {\textbar} {Transactions} of the {Association} for {Computational} {Linguistics} {\textbar} {MIT} {Press}},
	url = {https://direct.mit.edu/tacl/article/doi/10.1162/tacl_a_00471/110550/TopiOCQA-Open-domain-Conversational-Question},
	urldate = {2025-11-24},
}

@inproceedings{yang_adaptive_2024,
	address = {Washington DC USA},
	title = {On {Adaptive} {Knowledge} {Distillation} with {Generalized} {KL}-{Divergence} {Loss} for {Ranking} {Model} {Refinement}},
	isbn = {9798400706813},
	url = {https://dl.acm.org/doi/10.1145/3664190.3672522},
	doi = {10.1145/3664190.3672522},
	language = {en},
	urldate = {2025-11-24},
	booktitle = {Proceedings of the 2024 {ACM} {SIGIR} {International} {Conference} on {Theory} of {Information} {Retrieval}},
	publisher = {ACM},
	author = {Yang, Yingrui and He, Shanxiu and Yang, Tao},
	month = aug,
	year = {2024},
	pages = {81--90},
}

@inproceedings{yang_balanced_2023,
	address = {New York, NY, USA},
	series = {{ICTIR} '23},
	title = {Balanced {Knowledge} {Distillation} with {Contrastive} {Learning} for {Document} {Re}-ranking},
	isbn = {9798400700736},
	url = {https://dl.acm.org/doi/10.1145/3578337.3605120},
	doi = {10.1145/3578337.3605120},
	urldate = {2025-11-24},
	booktitle = {Proceedings of the 2023 {ACM} {SIGIR} {International} {Conference} on {Theory} of {Information} {Retrieval}},
	publisher = {Association for Computing Machinery},
	author = {Yang, Yingrui and He, Shanxiu and Qiao, Yifan and Xie, Wentai and Yang, Tao},
	month = aug,
	year = {2023},
	pages = {247--255},
}

@inproceedings{mo_aligning_2024,
	address = {New York, NY, USA},
	series = {{CIKM} '24},
	title = {Aligning {Query} {Representation} with {Rewritten} {Query} and {Relevance} {Judgments} in {Conversational} {Search}},
	isbn = {979-8-4007-0436-9},
	url = {https://dl.acm.org/doi/10.1145/3627673.3679534},
	doi = {10.1145/3627673.3679534},
	urldate = {2025-11-24},
	booktitle = {Proceedings of the 33rd {ACM} {International} {Conference} on {Information} and {Knowledge} {Management}},
	publisher = {Association for Computing Machinery},
	author = {Mo, Fengran and Qu, Chen and Mao, Kelong and Wu, Yihong and Su, Zhan and Huang, Kaiyu and Nie, Jian-Yun},
	month = oct,
	year = {2024},
	pages = {1700--1710},
}

@misc{arabzadeh_predicting_2021,
	title = {Predicting {Efficiency}/{Effectiveness} {Trade}-offs for {Dense} vs. {Sparse} {Retrieval} {Strategy} {Selection}},
	url = {http://arxiv.org/abs/2109.10739},
	doi = {10.48550/arXiv.2109.10739},
	urldate = {2025-11-24},
	publisher = {arXiv},
	author = {Arabzadeh, Negar and Yan, Xinyi and Clarke, Charles L. A.},
	month = sep,
	year = {2021},
	note = {arXiv:2109.10739 [cs]},
}

@article{kullback_information_1951,
	title = {On {Information} and {Sufficiency}},
	volume = {22},
	issn = {0003-4851, 2168-8990},
	url = {https://projecteuclid.org/journals/annals-of-mathematical-statistics/volume-22/issue-1/On-Information-and-Sufficiency/10.1214/aoms/1177729694.full},
	doi = {10.1214/aoms/1177729694},
	number = {1},
	urldate = {2025-11-04},
	journal = {The Annals of Mathematical Statistics},
	author = {Kullback, S. and Leibler, R. A.},
	month = mar,
	year = {1951},
	note = {Publisher: Institute of Mathematical Statistics},
	pages = {79--86},
}

@inproceedings{yu_few-shot_2021,
	address = {Virtual Event Canada},
	title = {Few-{Shot} {Conversational} {Dense} {Retrieval}},
	isbn = {978-1-4503-8037-9},
	url = {https://dl.acm.org/doi/10.1145/3404835.3462856},
	doi = {10.1145/3404835.3462856},
	language = {en},
	urldate = {2025-11-04},
	booktitle = {Proceedings of the 44th {International} {ACM} {SIGIR} {Conference} on {Research} and {Development} in {Information} {Retrieval}},
	publisher = {ACM},
	author = {Yu, Shi and Liu, Zhenghao and Xiong, Chenyan and Feng, Tao and Liu, Zhiyuan},
	month = jul,
	year = {2021},
	pages = {829--838},
}

@inproceedings{wu_conqrr_2022,
	address = {Abu Dhabi, United Arab Emirates},
	title = {{CONQRR}: {Conversational} {Query} {Rewriting} for {Retrieval} with {Reinforcement} {Learning}},
	shorttitle = {{CONQRR}},
	url = {https://aclanthology.org/2022.emnlp-main.679/},
	doi = {10.18653/v1/2022.emnlp-main.679},
	urldate = {2025-11-04},
	booktitle = {Proceedings of the 2022 {Conference} on {Empirical} {Methods} in {Natural} {Language} {Processing}},
	publisher = {Association for Computational Linguistics},
	author = {Wu, Zeqiu and Luan, Yi and Rashkin, Hannah and Reitter, David and Hajishirzi, Hannaneh and Ostendorf, Mari and Tomar, Gaurav Singh},
	editor = {Goldberg, Yoav and Kozareva, Zornitsa and Zhang, Yue},
	month = dec,
	year = {2022},
	pages = {10000--10014},
}

@inproceedings{mao_learning_2023,
	address = {Austin TX USA},
	title = {Learning {Denoised} and {Interpretable} {Session} {Representation} for {Conversational} {Search}},
	isbn = {978-1-4503-9416-1},
	url = {https://dl.acm.org/doi/10.1145/3543507.3583265},
	doi = {10.1145/3543507.3583265},
	language = {en},
	urldate = {2025-11-04},
	booktitle = {Proceedings of the {ACM} {Web} {Conference} 2023},
	publisher = {ACM},
	author = {Mao, Kelong and Qian, Hongjin and Mo, Fengran and Dou, Zhicheng and Liu, Bang and Cheng, Xiaohua and Cao, Zhao},
	month = apr,
	year = {2023},
	pages = {3193--3202},
}

@misc{hai_cosplade_2024,
	title = {{CoSPLADE}: {Contextualizing} {SPLADE} for {Conversational} {Information} {Retrieval}},
	shorttitle = {{CoSPLADE}},
	url = {http://arxiv.org/abs/2301.04413},
	doi = {10.48550/arXiv.2301.04413},
	urldate = {2025-11-04},
	publisher = {arXiv},
	author = {Hai, Nam Le and Gerald, Thomas and Formal, Thibault and Nie, Jian-Yun and Piwowarski, Benjamin and Soulier, Laure},
	month = jul,
	year = {2024},
	note = {arXiv:2301.04413 [cs]},
}

@article{mo_survey_2025,
	title = {A {Survey} of {Conversational} {Search}},
	volume = {43},
	issn = {1046-8188, 1558-2868},
	url = {https://dl.acm.org/doi/10.1145/3759453},
	doi = {10.1145/3759453},
	language = {en},
	number = {6},
	urldate = {2025-11-03},
	journal = {ACM Transactions on Information Systems},
	author = {Mo, Fengran and Mao, Kelong and Zhao, Ziliang and Qian, Hongjin and Chen, Haonan and Cheng, Yiruo and Li, Xiaoxi and Zhu, Yutao and Dou, Zhicheng and Nie, Jian-Yun},
	month = nov,
	year = {2025},
	pages = {1--50},
}

@inproceedings{xiao_beyond_2023,
	address = {Cham},
	title = {Beyond {Precision}: {A} {Study} on {Recall} of {Initial} {Retrieval} with {Neural} {Representations}},
	isbn = {978-3-031-24755-2},
	shorttitle = {Beyond {Precision}},
	doi = {10.1007/978-3-031-24755-2_7},
	language = {en},
	booktitle = {Information {Retrieval}},
	publisher = {Springer Nature Switzerland},
	author = {Xiao, Yan and Fan, Yixing and Zhang, Ruqing and Guo, Jiafeng},
	editor = {Chang, Yi and Zhu, Xiaofei},
	year = {2023},
	pages = {76--89},
}

@inproceedings{matveeva_high_2006,
	address = {Seattle Washington USA},
	title = {High accuracy retrieval with multiple nested ranker},
	isbn = {978-1-59593-369-0},
	url = {https://dl.acm.org/doi/10.1145/1148170.1148246},
	doi = {10.1145/1148170.1148246},
	language = {en},
	urldate = {2025-11-03},
	booktitle = {Proceedings of the 29th annual international {ACM} {SIGIR} conference on {Research} and development in information retrieval},
	publisher = {ACM},
	author = {Matveeva, Irina and Burges, Chris and Burkard, Timo and Laucius, Andy and Wong, Leon},
	month = aug,
	year = {2006},
	pages = {437--444},
}

@inproceedings{chen_efficient_2017,
	address = {Shinjuku Tokyo Japan},
	title = {Efficient {Cost}-{Aware} {Cascade} {Ranking} in {Multi}-{Stage} {Retrieval}},
	isbn = {978-1-4503-5022-8},
	url = {https://dl.acm.org/doi/10.1145/3077136.3080819},
	doi = {10.1145/3077136.3080819},
	language = {en},
	urldate = {2025-11-03},
	booktitle = {Proceedings of the 40th {International} {ACM} {SIGIR} {Conference} on {Research} and {Development} in {Information} {Retrieval}},
	publisher = {ACM},
	author = {Chen, Ruey-Cheng and Gallagher, Luke and Blanco, Roi and Culpepper, J. Shane},
	month = aug,
	year = {2017},
	pages = {445--454},
}

@inproceedings{formal_splade_2021,
	address = {New York, NY, USA},
	series = {{SIGIR} '21},
	title = {{SPLADE}: {Sparse} {Lexical} and {Expansion} {Model} for {First} {Stage} {Ranking}},
	isbn = {978-1-4503-8037-9},
	shorttitle = {{SPLADE}},
	url = {https://dl.acm.org/doi/10.1145/3404835.3463098},
	doi = {10.1145/3404835.3463098},
	urldate = {2025-11-03},
	booktitle = {Proceedings of the 44th {International} {ACM} {SIGIR} {Conference} on {Research} and {Development} in {Information} {Retrieval}},
	publisher = {Association for Computing Machinery},
	author = {Formal, Thibault and Piwowarski, Benjamin and Clinchant, Stéphane},
	month = jul,
	year = {2021},
	pages = {2288--2292},
}

@inproceedings{lupart_disco_2025,
	address = {Padua Italy},
	title = {{DiSCo}: {LLM} {Knowledge} {Distillation} for {Efficient} {Sparse} {Retrieval} in {Conversational} {Search}},
	isbn = {979-8-4007-1592-1},
	shorttitle = {{DiSCo}},
	url = {https://dl.acm.org/doi/10.1145/3726302.3729966},
	doi = {10.1145/3726302.3729966},
	language = {en},
	urldate = {2025-10-30},
	booktitle = {Proceedings of the 48th {International} {ACM} {SIGIR} {Conference} on {Research} and {Development} in {Information} {Retrieval}},
	publisher = {ACM},
	author = {Lupart, Simon and Aliannejadi, Mohammad and Kanoulas, Evangelos},
	month = jul,
	year = {2025},
	pages = {9--19},
}

@inproceedings{lassance_efficiency_2022,
	address = {Madrid Spain},
	title = {An {Efficiency} {Study} for {SPLADE} {Models}},
	isbn = {978-1-4503-8732-3},
	url = {https://dl.acm.org/doi/10.1145/3477495.3531833},
	doi = {10.1145/3477495.3531833},
	language = {en},
	urldate = {2025-11-03},
	booktitle = {Proceedings of the 45th {International} {ACM} {SIGIR} {Conference} on {Research} and {Development} in {Information} {Retrieval}},
	publisher = {ACM},
	author = {Lassance, Carlos and Clinchant, Stéphane},
	month = jul,
	year = {2022},
	pages = {2220--2226},
}

@inproceedings{nguyen2023unified,
  title={A unified framework for learned sparse retrieval},
  author={Nguyen, Thong and MacAvaney, Sean and Yates, Andrew},
  booktitle={European Conference on Information Retrieval},
  pages={101--116},
  year={2023},
  organization={Springer}
}

@article{lassance2024splade,
  title={SPLADE-v3: New baselines for SPLADE},
  author={Lassance, Carlos and D{\'e}jean, Herv{\'e} and Formal, Thibault and Clinchant, St{\'e}phane},
  journal={arXiv preprint arXiv:2403.06789},
  year={2024}
}

@article{oord2018representation,
  title={Representation learning with contrastive predictive coding},
  author={Oord, Aaron van den and Li, Yazhe and Vinyals, Oriol},
  journal={arXiv preprint arXiv:1807.03748},
  year={2018}
}

@inproceedings{10.1145/3269206.3271800,
author = {Zamani, Hamed and Dehghani, Mostafa and Croft, W. Bruce and Learned-Miller, Erik and Kamps, Jaap},
title = {From Neural Re-Ranking to Neural Ranking: Learning a Sparse Representation for Inverted Indexing},
year = {2018},
isbn = {9781450360142},
publisher = {Association for Computing Machinery},
address = {New York, NY, USA},
url = {https://doi.org/10.1145/3269206.3271800},
doi = {10.1145/3269206.3271800},
booktitle = {Proceedings of the 27th ACM International Conference on Information and Knowledge Management},
pages = {497–506},
numpages = {10},
location = {Torino, Italy},
series = {CIKM '18}
}

@article{paria2020minimizing,
  title={Minimizing flops to learn efficient sparse representations},
  author={Paria, Biswajit and Yeh, Chih-Kuan and Yen, Ian EH and Xu, Ning and Ravikumar, Pradeep and P{\'o}czos, Barnab{\'a}s},
  journal={arXiv preprint arXiv:2004.05665},
  year={2020}
}

@article{formal2021spladev2,
  title={SPLADE v2: Sparse lexical and expansion model for information retrieval},
  author={Formal, Thibault and Lassance, Carlos and Piwowarski, Benjamin and Clinchant, St{\'e}phane},
  journal={arXiv preprint arXiv:2109.10086},
  year={2021}
}

\clearpage
\onecolumn
\end{document}